\documentclass[sigconf]{acmart}
\usepackage[absolute,overlay]{textpos}
\usepackage{graphicx}
\usepackage{xcolor}
\usepackage{url}
\usepackage{subcaption}
\usepackage{graphicx}
\usepackage{xspace}
\usepackage{multirow}
\usepackage{comment}
\usepackage{color, colortbl}

\newcommand\blfootnote[1]{%
  \begingroup
  \renewcommand\thefootnote{}\footnote{#1}%
  \addtocounter{footnote}{-1}%
  \endgroup
}

\begin{document}

\title{Towards Understanding Political Interactions on Instagram}

\author{Martino Trevisan, Luca Vassio, Idilio Drago, Marco Mellia}
\affiliation{\institution{Politecnico di Torino}}

\author{Fabricio Murai, Flavio Figueiredo, Ana~Paula~Couto~da~Silva, Jussara M. Almeida}
\affiliation{\institution{Universidade Federal de Minas Gerais}}

\begin{abstract}

Online Social Networks (OSNs) allow personalities and companies to communicate directly with the public, bypassing filters of traditional medias. As people rely on OSNs to stay up-to-date, the political debate has moved online too. We witness the sudden explosion of harsh political debates and the dissemination of rumours in OSNs. Identifying such behaviour requires a deep understaning on how people interact via OSNs during political debates. We present a preliminary study of interactions in a popular OSN, namely Instagram. We take Italy as a case study in the period before the 2019 European Elections. We observe the activity of top Italian Instagram profiles in different categories: politics, music, sport and show. We record their posts for more than two months, tracking ``likes'' and comments from users. Results suggest that profiles of politicians attract markedly different interactions than other categories. People tend to comment more, with longer comments, debating for longer time, with a large number of replies, most of which are not explicitly solicited. Moreover, comments tend to come from a small group of very active users. 
Finally, we witness substantial differences when comparing profiles of different parties. 
\end{abstract}


% \copyrightyear{2019}
% \acmYear{2019}
% \acmConference[HT '19]{30th ACM Conference on Hypertext and Social Media}{September 17--20, 2019}{Hof, Germany}
% \acmBooktitle{30th ACM Conference on Hypertext and Social Media (HT '19), September 17--20, 2019, Hof, Germany}
% \acmPrice{15.00}
% \acmDOI{10.1145/3342220.3343657}
% \acmISBN{978-1-4503-6885-8/19/09}

\settopmatter{printacmref=false} 
\renewcommand\footnotetextcopyrightpermission[1]{}

\pagestyle{plain} % removes running headers

\maketitle

\TPshowboxestrue
\TPMargin{0.3cm}
\begin{textblock*}{18cm}(1.8cm,0.7cm)
\bf
\definecolor{myRed}{rgb}{0.55,0,0}
\color{myRed}
\noindent
Please cite this article as: Martino Trevisan, Luca Vassio, Idilio Drago, Marco Mellia, Fabricio Murai, Flavio Figueiredo, Ana Paula Couto da Silva, Jussara M. Almeida. Towards Understanding Political Interactions on Instagram. In Proceedings of the 30th ACM Conference on Hypertext and Social Media (pp. 247-251). 2019. DOI: \url{https://doi.org/10.1145/3342220.3343657}
\end{textblock*}

\section{Introduction}

%motivation & research questions
Online Social Networks (OSNs) have become a space of paramount importance for the exchange of content and dissemination of information. As more and more people rely on OSNs to stay up-to-date, the political debate has naturally moved online too. Politicians and political associations  communicate directly with the public via OSNs, bypassing filters of journalists and traditional media. While this lack of mediation has opened unprecedented forms of interactions, we witness the explosion of harsh political debates, possibly fuelled by provoking users, as well as the dissemination of rumours and hate speech in OSNs. Identifying and fighting such online toxic behaviour is a major open problem, and requires a deep understanding on how people face the political debate via OSNs. 

% what we have done
In this paper, we aim at understanding the peculiarities of political interactions on Instagram, an OSN that is more and more prominent among users~\cite{Trevisan:2018}. In this preliminary study, we focus on the activity of top public Instagram profiles (nowadays known as ``influencers'') in Italy before the European Elections of May 2019. We take a close look into the activity of politicians and the users' interactions across different political groups. To understand whether interactions across political figures follow general patterns, we also track the activity of public figures in other categories, namely music, sport and show/TV/entertainment. We register their ``posts'' for two months, tracking ``likes'' and comments from users. We then investigate whether these interactions are similar across profiles of different types of influencers. 

% Results highlights
Our preliminary results suggest that interactions across types of profiles are largely dissimilar. In particular, profiles of politicians appear to attract peculiar interaction types. First, politicians' posts attract more and lengthier comments, for longer time periods. Second, users hardly mention other users when commenting politicians' posts, whereas mentions are fairly common in posts of other types of influencers. Third, comments in politicians' posts attract a large number of replies (i.e., comments to previous comments) and most of them are unsolicited, a fact which is generally unusual. At last, comments in politicians' posts tend to come from a small group of very active users. This behaviour suggests the existence of a group of users that actively participate in discussions and reply to other comments possibly aiming to influence the online political debate. Interestingly, we also witness substantial differences among political parties. 

% Related work on politics on social networks
Some recent works have focused on how users interact with social networks and political content. Howard \emph{et al.}~\cite{howard2016bots} study the role of bots on political conversations on Twitter, while Mahoney \emph{et al.}~\cite{mahoney2016constructing} investigate how the Scottish electorate utilises Instagram for political self-expression. 
Resende \emph{et al.}~\cite{Whatsapp19} analyse information dissemination within WhatsApp political-oriented groups, collecting all messages shared during the Brazilian presidential campaign in 2018. 
Instagram has already been proven useful to study human behaviour on a variety of fields, from funerary practices~\cite{gibbs2015funeral}, to depression~\cite{reece2017instagram}, body image satisfaction~\cite{ridgway2016instagram, sheldon2016instagram},  photojournalism~\cite{borges2015news}. Hu \emph{et al.}~\cite{hu2014we} use computer vision techniques to study Instagram users, while Highfield \emph{et al.}~\cite{highfield2015methodology} focus on hashtags, and compare their usage with Twitter. Our work instead focuses on how people interact with politicians and personalities on Instagram.

\blfootnote{The research leading to these results has been funded by the foundation Compagnia di San Paolo through the 2019 internationalisation project program between Politecnico di Torino (Italy) and Universidade Federal de Minas Gerais (Brazil), by the Brazilian research agencies CNPq, CAPES and FAPEMIG and by the SmartData@Polito center.}

\section{Methodology}

This section describes how we collect Instagram data and provides some definitions of entities used  to characterise interactions on this social network.

\subsection{Data collection}
\label{sec:dataset}

We focus on data collected from Instagram \emph{public profiles}. We use a custom crawler to download and store data and meta-data regarding the profiles and the corresponding posts and comments.

Our crawler collects the activity of the set of monitored profiles in real-time. Periodically, it downloads the meta-data of the profiles and all their new generated content, i.e., their posts. For these posts, our crawler downloads all the comments written by any user in the first 24 hours after the posting time. We remove all sensitive information from the data, e.g., any account identification of the users, and store the remaining information on a Hadoop-based cluster for further processing.

We are interested in the \emph{top public figures} that publish information via Instagram. As such, we set an arbitrary threshold to include a profile in the crawling: Only profiles with at least $10\,000$ followers are considered. Since here we evaluate Italian profiles only, we further restrict the data capture to profiles whose posts are composed by at least $40\%$ of Italian words.\footnote{We use a Hunspell dictionary to filter Italian words. It is available at: \url{https://cgit.freedesktop.org/libreoffice/dictionaries/tree/}} 

The list of monitored profiles is not fixed and has grown during the crawling campaign. We search for new profiles using the \emph{hashtags} present in posts, and whenever a profile explicitly mentions another one. We started the crawler on late December 2018, with a list of $50$ popular Italian public figures. The profile list started growing very fast, and after two weeks it already included more than $10\,000$ entries. After four months of collection (April 2019), the crawler includes $19\,156$ Italian public profiles that generated $1\,367\,949$ posts associated to $57\,617\,533$ comments. To avoid possible artefacts due to the fact that some profiles have been added to the crawler after than others, for the analysis that follows we consider only profiles present in the crawler on February 1st, 2019. Moreover, we evaluate only posts and comments from February 1st to April 10th, 2019.

Instagram provides little information about a profile: a name, a biography and a profile picture. We thus resort to external sources to map each profile to a category and each politician's profile to a political party. First, we use HypeAuditor,\footnote{\url{https://hypeauditor.com/}} an online analytics platform, to get the list of top Italian influencers. HypeAuditor's public list is restricted to the top-1000 profiles per country, from which we obtain $150$ profiles belonging to the categories of our interest (i.e., sport, music and show) and passing the filters described above (e.g., posting mostly in Italian in the evaluated time period). Second, to find politicians, we search the set of $19\,156$ monitored profiles, retrieving $85$ profiles belonging to five Italian major political parties, grouped into three factions:\footnote{https://en.wikipedia.org/wiki/List\_of\_political\_parties\_in\_Italy} (i) Lega Nord (Lega), Forza Italia (FI) and Fratelli d'Italia (FdI), which are centre-right parties,\footnote{Nevertheless, these three parties run autonomously for the 2019 European Election.} (ii) Partito Democratico (PD), the main progressive party, (iii) Movimento Cinque Stelle (M5S), the main independent anti-establishment party. In the remainder of the paper we limit our analysis to those categorised profiles, summarised in Table~\ref{tab:dataset}.

\begin{table}[t]
    \caption{Dataset overview. The last three rows are subclasses of politics.}
    \begin{tabular}{l|r|r|r|r}
    Class        & Profiles    & Posts   & Comments    & Commenters  \\ \hline
    Music        & 50          & 2\,846  & 1\,565\,141 & 653\,683    \\
    Sport        & 68          & 4\,393  & 1\,773\,835 & 724\,959    \\
    Show         & 32          & 2\,338  & 1\,183\,989 & 565\,126    \\
    Politics     & 85          & 7\,617  & 1\,621\,857 & 306\,541    \\ \hline
    Lega + FI + FdI  & 32          & 4\,107  & 1\,225\,597 & 234\,745    \\
    M5S          & 33          & 2\,037  & 276\,593    & 70\,075     \\
    PD           & 20          & 1\,473  & 119\,667    & 33\,647     \\
    \end{tabular}
    \label{tab:dataset}
\end{table}
\subsection{Categorising mentions}
\label{sec:methodology}

\begin{figure}[t]
    \begin{center}
        \includegraphics[width=\columnwidth]{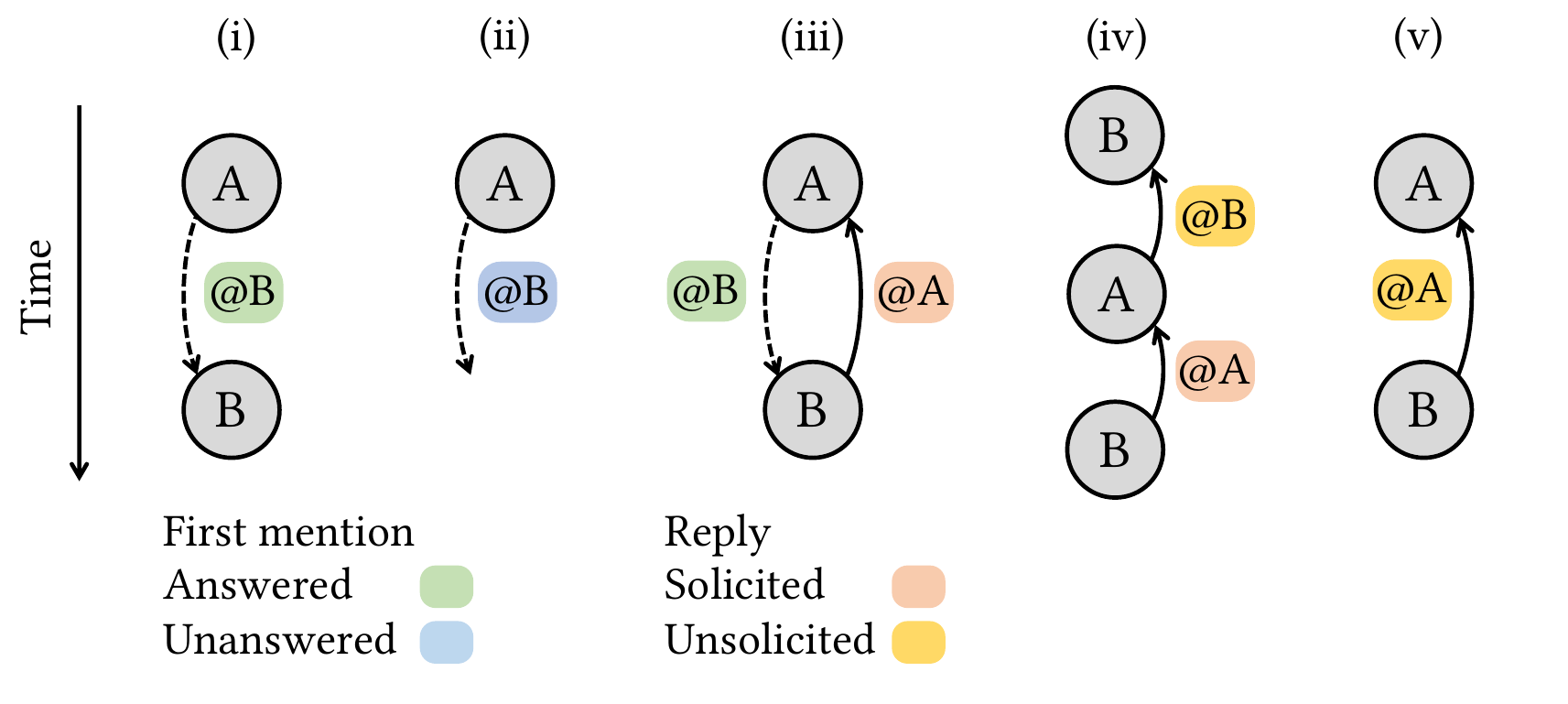}
        \caption{Examples of mentions: Users' comments are represented as circles, mentions as arrows. Time evolves top-down. Colours represent the way we categorise mentions.}
        \label{fig:mention_examples}
    \end{center}
\end{figure}

Each post may receive many comments. We call a user commenting in a post a ``commenter''. 
Then, we are interested in characterising how users react to ``mentions'' -- i.e., the special case of comments containing explicit references to other users. Mentions may represent different situations, such as users calling the attention of others to a discussion or users answering comments of others. 

We classify the mentions by evaluating the comments of a post chronologically, from the oldest to the newest comment, and checking the mentioning/mentioned users. We classify all the mentions into four categories which are exemplified with colours in Figure~\ref{fig:mention_examples}. 

First, whenever we encounter a comment from user \texttt{A} mentioning user \texttt{B}, who has not commented on the post yet, we classify this \emph{first mention} found in \texttt{A}'s comment as:\footnote{Note that we may miscount some of these two cases due to border effects, since we stop monitoring the comments of a post after 24 hours.}
\begin{itemize}
    \item {\it Answered first mention}: If \texttt{B} comments at least once on the post after \texttt{A}'s comment. Note that we consider the mention in \texttt{A}'s comment as answered if \texttt{B} comments afterwards, regardless of whether \texttt{B}'s comment contains other mentions;
    \item {\it Unanswered first mention}: If \texttt{B} never comments on the post after the mention of user \texttt{A}. 
\end{itemize}
Here we are interested in evaluating the effectiveness of explicitly calling someone else to the debate of a post. In this case, we consider that user \texttt{A} is calling user \texttt{B} to the debate. These two cases are illustrated in Figure~\ref{fig:mention_examples}~(i) and~(ii).

Second, whenever we find a comment of a user \texttt{B} mentioning user \texttt{A}, who has already commented on the post, we classify the mention found in \texttt{B}'s comment as:  
\begin{itemize}
    \item {\it Solicited reply}: If \texttt{A} has mentioned \texttt{B} before in the post;
    \item {\it Unsolicited reply}: if \texttt{A} has not mentioned \texttt{B} before in the post.
\end{itemize}
In these cases we evaluate whether people engage in conversations and whether people reply to comments without being invited to the debate. These cases are illustrated in Figure~\ref{fig:mention_examples}~(iii)--(v). Note that a mention classified as ``solicited reply'' may appear after another one classified as ``answered first mention'' (Figure~\ref{fig:mention_examples}~(iii)) as well as after a mention classified as ``unsolicited reply'' (Figure~\ref{fig:mention_examples}~(iv)).

It is important to highlight a difference between the above definitions and the representation of comments and replies in Instagram's apps. When listing comments and replies, Instagram not only shows mentions between users, but also associates the precise comment a user is referring to when mentioning someone. Our crawler cannot pinpoint these associations, receiving instead a list of comments and mentions in chronological order. Due to this limitation, we only focus on counting and classifying mentions as described above.

\section{Results}
\label{sec:results}

In this section we start analysing the number of comments and how they are distributed across commenters. We then focus on the comment length and the speed at which they are created after a post. Finally we quantify interactions among commenters analysing mentions. 

\begin{figure}[t]
    \begin{center}
        \begin{subfigure}{0.45\textwidth}
            \includegraphics[width=\columnwidth]{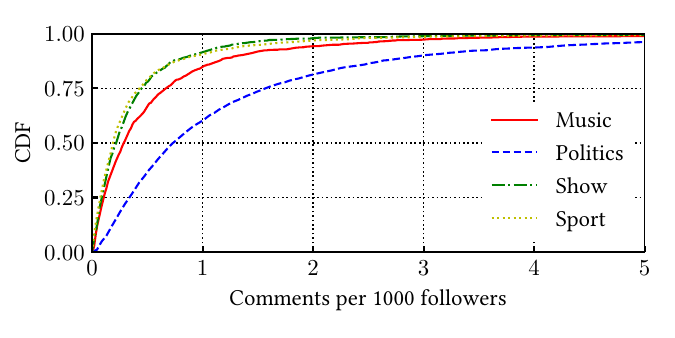}
            \vspace{-5mm}
            \caption{All categories}
            \label{fig:comment_number_all}
        \end{subfigure}
        \begin{subfigure}{0.45\textwidth}
            \includegraphics[width=\columnwidth]{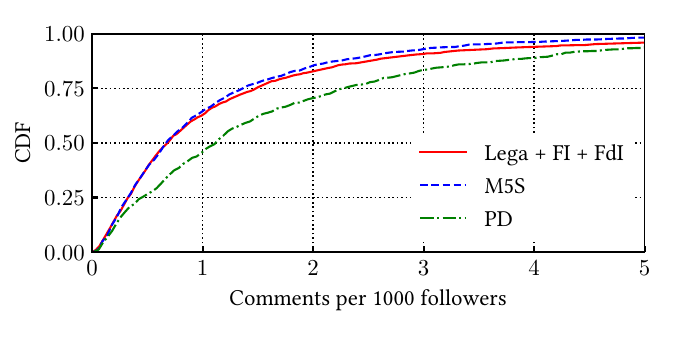}
            \vspace{-5mm}
            \caption{Politics}
            \label{fig:comment_number_politics}
        \end{subfigure}
        \vspace{-2mm}
        \caption{CDFs of number of comments per post normalised by 1\,000 followers.}
        \vspace{-2mm}
        \label{fig:comment_number}
    \end{center}
\end{figure}

\begin{figure}[t]
    \begin{center}
        \begin{subfigure}{0.45\textwidth}
            \includegraphics[width=\columnwidth]{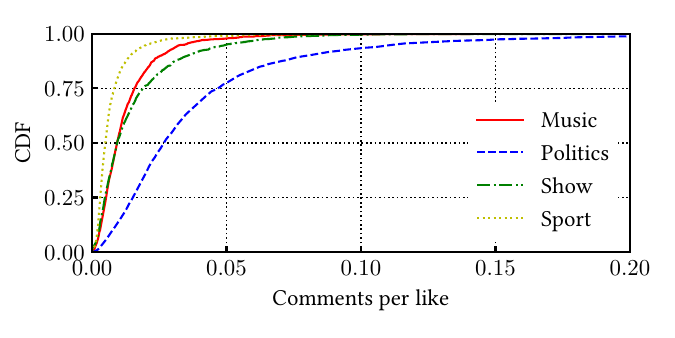}
            \vspace{-5mm}
            \caption{All categories}
            \label{fig:comment_number_vs_like_all}
        \end{subfigure}
        \begin{subfigure}{0.45\textwidth}
            \includegraphics[width=\columnwidth]{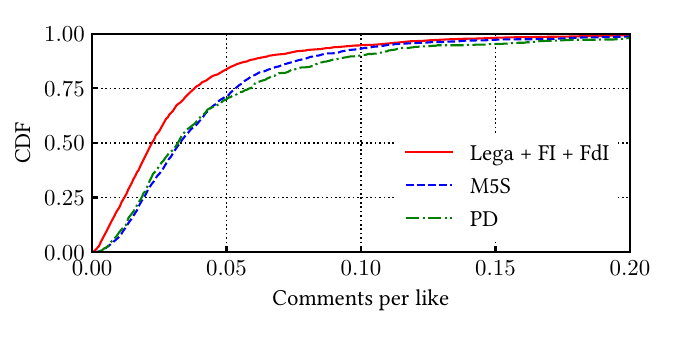}
            \vspace{-5mm}
            \caption{Politics}
            \label{fig:comment_number_vs_like_politics}
        \end{subfigure}
        \vspace{-2mm}
        \caption{CDFs of ratio between number of comments and number of likes per post.}
        \vspace{-2mm}
        \label{fig:comment_number_vs_like}
    \end{center}
\end{figure}

\subsection{Volume of comments}
\label{sec:volume_comments}

We first quantify the number of comments received by influencers, grouped by category. Figure~\ref{fig:comment_number_all} shows the empirical Cumulative Distribution Functions (CDFs) of the number of comments per post normalised by the number of followers of the influencers. Normalisation is necessary due to high correlation between these two quantities. Different curves refer to different categories. Overall, the number of comments by 1\,000 followers is quite low, with only few posts collecting more than 5 comments every 1\,000 followers. Sport and show exhibit quite similar behaviour, with music having a relatively larger number of comments per follower. Politics instead shows a remarkable difference: the normalised median number of comments per post is 3.2 times larger than for sport with 8\% of posts collecting more than 4 comments per 1\,000 follower. Figure~\ref{fig:comment_number_politics} breaks down the analysis for political parties. Posts of PD receive more comments: its median is 1.1 comments every 1\,000 followers while for other parties it is less than 0.7.

We next compare the number of comments with respect to the number of likes per post. Figure~\ref{fig:comment_number_vs_like} shows the CDFs of the ratio between the number of comments and the number of likes. We observe that posts tend to collect many more likes than comments. This low ratio of comments per like is expected, and reflects the fact that most Instagram users limit their interactions to a like, with only few users actively commenting on posts. Yet, for politics, comments are sensibly more frequent than for other categories. Considering political parties, we observe that posts from centre-right parties have a smaller ratio of comments per like when compared to other groups. 

\begin{figure}[!t]
    \begin{center}
        \includegraphics[width=0.90\columnwidth]{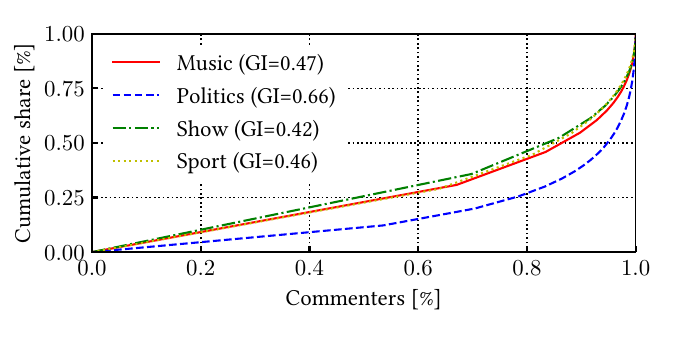}
        \vspace{-5mm}
        \caption{Lorenz curve for the number of comments per commenter. Gini Index is reported in the legend.}
        \vspace{-2mm}
        \label{fig:comment_lorenz}
    \end{center}
\end{figure}

Let us focus on how concentrated the comments are with respect to the population of commenters. We use the Lorenz curve for this, depicted in Figure~\ref{fig:comment_lorenz}. Each curve shows the fraction of comments written by the bottom $x\%$ of the commenters (sorted by this quantity). 
The more the Lorenz curve leans towards the right bottom of the plot, the more inequalities are present. Consider the area between the line of perfect equality (the main diagonal) and the Lorenz curve. Consider then the area between the line of perfect equality and the line of perfect inequality (the x axis). The ratio between these two areas is the Gini Index (GI). The closer to 1 the index is, the more unequal the distribution is.

Results are striking: the inequality for commenters of politics is much larger than the ones in other categories (Gini Index is 0.66 versus 0.42-0.47). For instance, 20\% of the commenters are responsible for 75\% of all comments in politics. This result hints for a small group of very active commenters in politics, not seen in other categories. Indeed, the top-100 commenters in politics are responsible for more than $40 k$ comments ($2.5\%$ of the total for politics), while the top-100 commenters for other categories are responsible for less than $20 k$ ($0.4\%$ of the total for the other categories).
The breakdown across political parties, omitted for the sake of brevity, does not highlight major differences among them. The Gini Index is $0.62$ for M5S and PD and $0.65$ for conservative parties (Lega + FI + FdI).

\subsection{Comment length and timing}

\begin{figure}[t]
    \begin{center}
        \begin{subfigure}{0.45\textwidth}
            \includegraphics[width=\columnwidth]{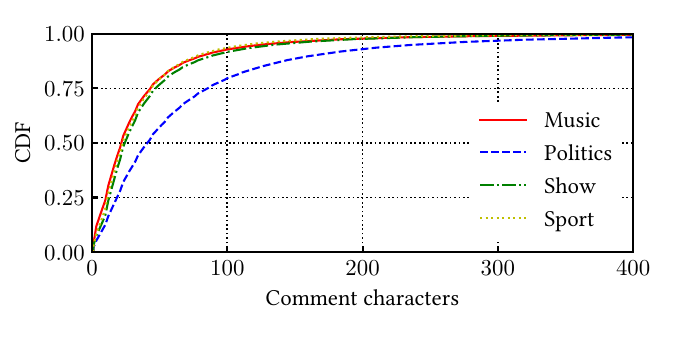}
            \vspace{-5mm}
            \caption{All categories}
            \label{fig:comment_len_all}
        \end{subfigure}
        \begin{subfigure}{0.45\textwidth}
            \includegraphics[width=\columnwidth]{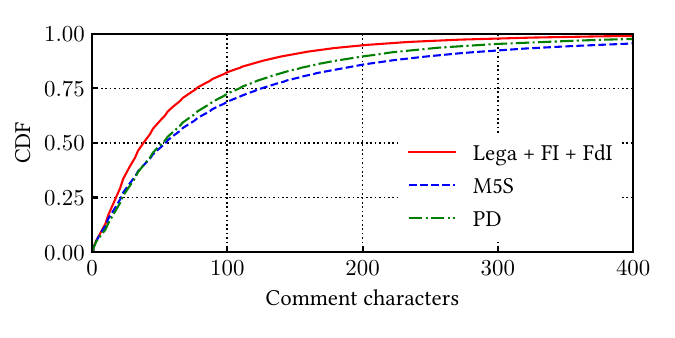}
            \vspace{-5mm}
            \caption{Politics}
            \label{fig:comment_len_politics}
        \end{subfigure}
        \vspace{-2mm}
        \caption{CDFs of number of characters per comment.}
        \vspace{-2mm}
        \label{fig:comment_len}
    \end{center}
\end{figure}

We now analyse the length of comments and show in Figure~\ref{fig:comment_len_all} the CDFs of the number of characters per comment. Most comments are quite short, i.e., 75\% are shorter than 50 characters for music, show and sport. However, politics collects much longer comments, with the 75-percentile reaching 83 characters, and the 90-percentile 4 times larger than for other categories. Indeed, only 7\% of them are composed by a single \emph{emoji}, while this happens in 8-11\% of cases for other categories. Figure~\ref{fig:comment_len_politics} shows a breakdown by political faction. We notice that posts of conservative parties tend to collect shorter comments than PD and M5S.

\begin{figure}[t]
    \begin{center}
        \begin{subfigure}{0.45\textwidth}
            \includegraphics[width=\columnwidth]{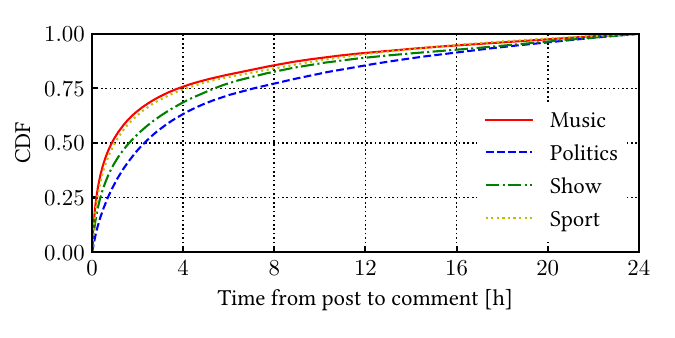}
            \vspace{-5mm}
            \caption{All categories}
            \label{fig:comment_delta_all}
        \end{subfigure}
        \begin{subfigure}{0.45\textwidth}
            \includegraphics[width=\columnwidth]{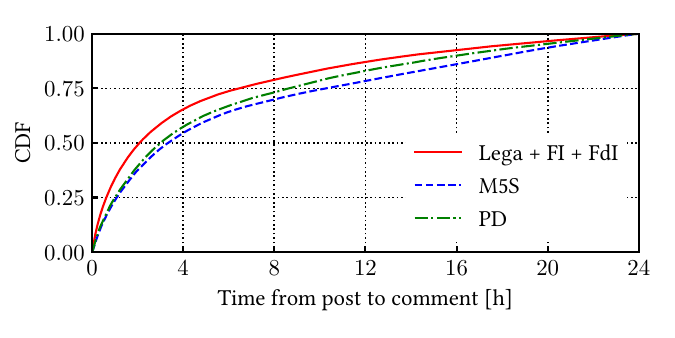}
            \vspace{-5mm}
            \caption{Politics}
            \label{fig:comment_delta_politics}
        \end{subfigure}
        \vspace{-2mm}
        \caption{CDFs of time interval between a comment and corresponding post (scale in hours).}
        \vspace{-2mm}
        \label{fig:comment_delta}
    \end{center}
\end{figure}

Let us now focus on the longevity of the interactions with a post. We measure it analysing the time passed from the moment a post is published to the time users post comments on it. This helps understanding how long users interact with a post -- i.e., if comments appear immediately after the post or spread in time. Figure~\ref{fig:comment_delta_all} shows the CDF of this time interval for different categories. Results show that music and sport collect the majority of comments close to the instant of publication. In contrast, show and especially politics collect larger fractions of comments further away in time. For instance, 75\% of comments to politics posts are made within 8 hours after being posted, while, for music, this happens within 4 hours. This may hint that political discussion and debate last longer with respect to other topics where discussions among users quickly fade away. In the next paragraph we will analyse this aspect in detail. Considering the different political parties (Figure~\ref{fig:comment_delta_politics}), conservative parties comments end sensibly in shorter time.

\subsection{Mentions in comments}

To further examine how users' interactions are generated, we study mentions between them. Intuitively, mentions can be an effective way of raising the number of comments in a post.

\begin{figure}[t!]
    \begin{center}
        \includegraphics[width=0.9\columnwidth]{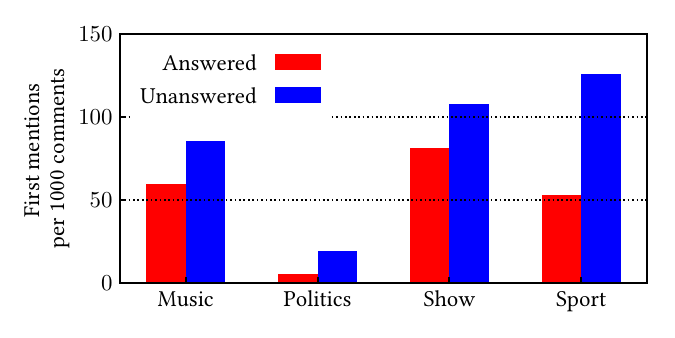}
        \vspace{-5mm}
        \caption{Number of answered and unanswered first mentions every 1\,000 comments.}
        \vspace{-2mm}
        \label{fig:mentions}
    \end{center}
\end{figure}

In Figure~\ref{fig:mentions} we report the number of first mentions per 1\,000 comments. Recall that a first mention is a mention toward a user that have not yet commented on this post. We distinguish between first mentions that are answered (red bars) and those that are left unanswered (blue bars). Once again, politics show a remarkably different behaviour than other categories: Every 1\,000 comments, there are only 24 first mentions,\footnote{A comment may contain more than one mention. Hence, the maximum number of comments including first mentions is 24 every 1\,000 comments.} and only 5 (20\%) get answered. For comparison, for show category, there are 188 mentions in 1\,000 comments, from which about 43\% get answered. In a sense, this could point out that most initial mentions in politics posts are not focused on attract new users to the discussion.

\begin{figure}[t]
    \begin{center}
        \includegraphics[width=0.9\columnwidth]{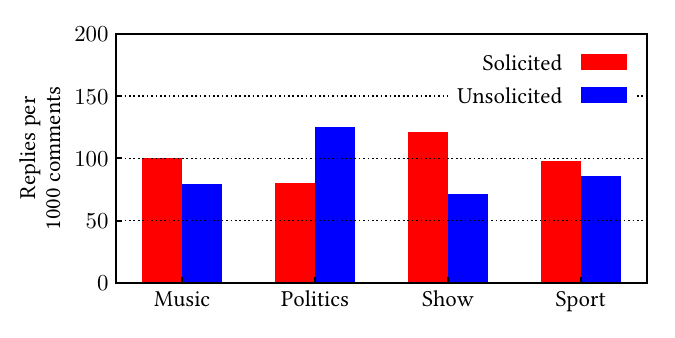}
        %\vspace{-5mm}
        \caption{Number of solicited and unsolicited replies every 1\,000 comments.}
       %\vspace{-2mm}
        \label{fig:reply}
    \end{center}
\end{figure}

The behaviour of users is even more interesting when we look at replies. Recall that a reply is a comment that mentions a user that has also previously commented. Replies help us to understand how users commenting on a post interact with each other. Figure~\ref{fig:reply} shows the number of replies every 1\,000 comments, for solicited (red bars) and unsolicited (blue bars) replies. Here, politics exhibits the largest number of replies 205, most of which are unsolicited. This means that users reply to comments of previous users, without being explicitly consulted. Analysing in depth the content of such replies as well as the users generating them is left for future work.

Overall, when we combine both results, it appears that political comments do not try to drag other users into the discussion (Figure~\ref{fig:mentions}), but instead they tend to reply to prior comments (Figure~\ref{fig:reply}). This may indicate that users start discussions after reading the opinion of others, possibly even without knowing each other. If we put these numbers in perspective with the fact that comments in politics are more concentrated around a small fraction of very active users (Section~\ref{sec:volume_comments}), these results may also be interpreted as an indication of the existence of a passionate group of users, actively reacting to comments, e.g., to influence the political debate. 

\section{Conclusion}

We analysed how users interact with politicians and other influencers on Instagram. Based on a large dataset including hundreds of Italian public profiles and millions of comments, we find notable differences across categories. 
Comments to politicians are more frequent and longer in comparison to other categories. Political commenters tend to discuss more, but are less likely to drag users not yet engaged in the discussion.
This work aims at fostering further researches in this field. We are currently monitoring the evolution in time of the interactions before and after the 2019 European Parliament election.\footnote{Daily updated results are available at \url{https://smartdata.polito.it/instagram_monitoring/}}

%we plan to monitor user interactions in the periods before and after the 2019 European Parliament election. In particular, we are working to extend our analysis to include other countries of the European Union.

\bibliographystyle{ACM-Reference-Format}
\bibliography{main}

\end{document}